\documentclass[reprint,
superscriptaddress, 
amsmath,
amssymb, 
aps,
floatfix
 ]{revtex4-2}
\usepackage{xcolor}
\usepackage{verbatim}
\usepackage{graphicx}
\usepackage{dcolumn}
\usepackage{bm}
\usepackage{upgreek}
\usepackage{array}
\usepackage{hyperref}
\usepackage{amsmath}
\begin{document}

\title{Mobility exceeding 100,000~cm$^2$/Vs in modulation-doped shallow InAs quantum wells coupled to epitaxial aluminum}

\author{Teng Zhang}
\affiliation{Department of Physics and Astronomy, Purdue University, West Lafayette, Indiana 47907, USA}
\affiliation{Birck Nanotechnology Center, Purdue University, West Lafayette, Indiana 47907, USA}

\author{Tyler Lindemann}
\affiliation{Department of Physics and Astronomy, Purdue University, West Lafayette, Indiana 47907, USA}%
\affiliation{Birck Nanotechnology Center, Purdue University, West Lafayette, Indiana 47907, USA}
\affiliation{Microsoft Quantum Lab West Lafayette, West Lafayette, Indiana 47907, USA}

\author{Geoffrey C. Gardner}
\affiliation{Birck Nanotechnology Center, Purdue University, West Lafayette, Indiana 47907, USA}
\affiliation{Microsoft Quantum Lab West Lafayette, West Lafayette, Indiana 47907, USA}

\author{Sergei Gronin}
\affiliation{Birck Nanotechnology Center, Purdue University, West Lafayette, Indiana 47907, USA}
\affiliation{Microsoft Quantum Lab West Lafayette, West Lafayette, Indiana 47907, USA}

\author{Tailung Wu}
\affiliation{Birck Nanotechnology Center, Purdue University, West Lafayette, Indiana 47907, USA}
\affiliation{Microsoft Quantum Lab West Lafayette, West Lafayette, Indiana 47907, USA}
 
\author{Michael J. Manfra}
\email{mmanfra@purdue.edu}
\affiliation{Department of Physics and Astronomy, Purdue University, West Lafayette, Indiana 47907, USA}
\affiliation{Birck Nanotechnology Center, Purdue University, West Lafayette, Indiana 47907, USA}
\affiliation{Microsoft Quantum Lab West Lafayette, West Lafayette, Indiana 47907, USA}
\affiliation{School of Materials Engineering, Purdue University, West Lafayette, Indiana 47907, USA}
\affiliation{Elmore Family School of Electrical and Computer Engineering, Purdue University, West Lafayette, Indiana 47907, USA}

\date{\today}
\begin{abstract}
 The two-dimensional electron gas residing in shallow InAs quantum wells coupled to epitaxial aluminum is a widely utilized platform for exploration of topological superconductivity. Strong spin-orbit coupling, large effective $g$-factor, and control over proximity-induced superconductivity are important attributes. Disorder in shallow semiconductor structures plays a crucial role for the stability of putative topological phases in hybrid structures. We report on the transport properties of 2DEGs residing 10nm below the surface in shallow InAs quantum wells in which mobility may exceed 100,000 cm$^2$/Vs at 2DEG density n$_{2DEG}$$\leq$1$\times$10$^{12}$cm$^{-2}$ at low temperature.
\end{abstract}

\maketitle

\section{Introduction}
Epitaxial semiconductor-superconductor hybrid materials provide a platform for exploring topological superconductivity ~\cite{Lutchyn.2010, Oreg.2010, Wan.2015, Chang.2015, Krogstrup.2015, Shabani.2016, Kjaergaard.2016, Pientka.2017, Nichele.2017,Karzig.2017, Drachmann.2017, Fornieri.2019, Ren.2019, Mayer.2020, Dartiailh.2021, Dartiailh.2021an9, Kanne.2021, Drachmann.2021, Banerjee.2022}. Among the various superconductor-semiconductor combinations, the two-dimensional electron gas (2DEG) in InAs quantum wells separated from epitaxially-grown aluminum by a thin InGaAs barrier has yielded significant results ~\cite{Nichele.2017, Fornieri.2019, Dartiailh.2021, Dartiailh.2021an9, Banerjee.2022}. Large effective $g$-factor, strong spin-orbit coupling, and controlled proximity coupling are useful properties of this system. Simultaneously, low-disorder materials are desirable to promote strong correlations~\cite{Pan.2020, Ahn.2021, Tian.2021}. The highest reported peak 2DEG mobility in an InAs quantum well grown on InP exceeds $10^6$~cm$^2$/Vs~\cite{Hatke.2017}. However, these heterostructures utilized deep quantum wells with the 2DEG residing 100~nm below the top surface, making them unwieldy for induced superconductivity experiments~\cite{Hatke.2017}. To control proximity coupling between the InAs 2DEG and aluminum, a thin top barrier ($\sim$10~nm) between the quantum well and the superconductor is usually employed. Due to the thinnest of this top barrier, the InAs 2DEG is sensitive to surface scattering in areas not covered by aluminum. Some experiments have suggested that nanofabrication can increase surface scattering~\cite{Drachmann.2021, Pauka.2020} while theory suggests high mobility is necessary to support topological phases under realistic device conditions \cite{Ahn.2021, topogap}. It is therefore desirable to explore approaches to enhance mobility in near surface 2DEGs while maintaining strong spin-orbit coupling and ease of inducement of superconductivity.

In this study, we compare undoped and silicon (Si) $\delta$-doped near-surface InAs/InGaAs heterostructures coupled to epitaxial aluminum grown on InP substrates. We systematically vary the position and density of the silicon donor layer to study the impact on the electronic properties of the 2DEG including mobility and spin-orbit coupling. The peak mobilities of optimized samples exceeds 100,000~cm$^2$/V~s at n$_{2DEG}$$\leq$10$^{12}$cm$^{-2}$ in gated Hall bar devices while similarly constructed undoped structures have peak mobility of approximately 57,000~cm$^2$/Vs at n$_{2DEG}$=6$\times$10$^{11}$cm$^{-2}$. Additionally, spin-orbit coupling strength as determined by weak antilocalization (WAL) analysis for Si $\delta$-doped samples reveals a non-monotonic relationship with 2DEG density. The induced superconducting gap of samples with epitaxial aluminum were determined using tunneling spectroscopy measurements; our analysis indicates that there is no substantial difference in the proximity effect between doped and undoped samples.

\section{MBE growth, device fabrication, and measurements}
\begin{figure}
\includegraphics{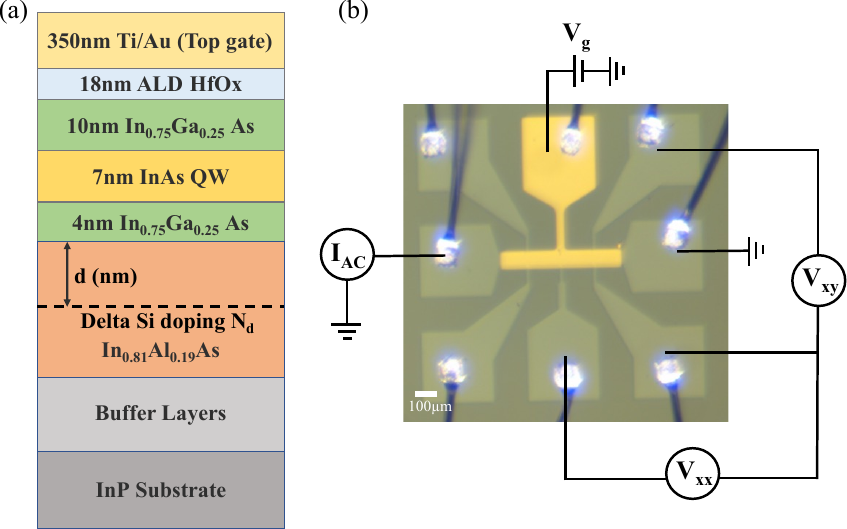}
\caption{\label{fig_1} (a) Layer stack of Samples A-F. The epitaxial Al layer is not shown here. $d$ is the spacer thickness, and $N_{d}$ is the doping density in the Si $\delta$-doping layer (black dashed line). (b) Optical image of a Hall bar and schematic of the experimental setup for magnetotransport measurements. The distance between two nearest voltage probes is $L$=100~\textmu{}m, and the width of the Hall bar is $W$=40~\textmu{}m. Black lines in the optical image are shadows of bonding wires.
}
\end{figure}

Six wafers with various doping parameters were grown via molecular beam epitaxy in a Veeco GEN 930 system. The sample parameters are detailed in Table ~\ref{Table1}, and a diagram of the semiconductor layer stack is shown in Fig.~\ref{fig_1}(a). Surface reconstruction during growth is monitored via reflection high-energy electron diffractometry (RHEED), and substrate temperature is measured via a thermocouple behind the substrate and via optical pyrometry. Semi-insulating Fe-doped InP (001) substrates were used for all growths in this study. The MBE chamber and growth materials were prepared as described by Gardner et al.~\cite{GARDNER201671}.

The native oxide of the InP substrate is thermally desorbed under As$_{4}$ overpressure, which is maintained at constant value of 10$^{-5}$~Torr throughout the growth. The oxide desorption occurs when the surface switches to a metal-rich 4$\times$2 reconstruction, after which the sample is immediately cooled and a 2$\times$4 reconstruction is recovered. After oxide desorption, a layer of lattice-matched InAlAs is grown to smooth the sample surface. InAs will relax if grown directly on an InP substrate due to the 3.3\% lattice mismatch between the two materials. Therefore, an InAlAs graded buffer layer (GBL) is used in order to provide a pseudosubstrate with a lattice constant closer to that of InAs.

Upon completion of the GBL, the temperature of the substrate is set to 480$^{\circ}$C as determined by pyrometry. A 58~nm In$_{0.81}$Al$_{0.19}$As layer is grown between the GBL and the lower InGaAs barrier. For doped samples, a Si-doping layer interrupts the 58~nm InAlAs layer, according to desired Si-doping density and setback. The active region consists of a 4~nm In$_{0.75}$Ga$_{0.25}$As lower barrier, a 7~nm InAs quantum well, and a 10~nm In$_{0.75}$Ga$_{0.25}$As top barrier. Following completion of the semiconductor growth, the sample is cooled and 5~nm of Al is epitaxially deposited on the InGaAs top barrier.

\begin{table*}
\caption{\label{Table1}%
Sample parameters including 2D doping density ($N_{d}$), spacer thickness ($d$), peak mobility ($\mu_{max}$), and 2DEG density at the peak mobility ($n_{2DEG}$ at $\mu_{max}$). Sample A is undoped. The dielectric material for all samples in this table is hafnium oxide.}
\begin{ruledtabular}
\begin{tabular}{c|c|c|c|c|c} 
 Sample & $N_d$ (10$^{12}$~cm$^{-2}$) & $d$ (nm) & $\mu_{max}$ ($10^3$~cm$^2$/Vs) & $n_{2DEG}$ at $\mu_{max}$ ($10^{12}$~cm$^{-2}$) & $n_{2DEG}$ at $V_{g}$ = 0 V ($10^{12}$~cm$^{-2}$)\\
 \hline
 A & N.A. & N.A. & 57 & 0.6 & 2.7\\ 
 B & 0.8 & 15 & 100 & 0.7 & 2.9\\ 
 C & 1 & 15 & 90  & 0.8 & 3.4 \\ 
 D & 2 & 15 & 120 & 0.9 & 3.3 \\ 
 E & 1 & 30 & 55 & 0.8 & 2.9\\ 
 F & 2 & 30 & 65 & 0.8 & 3.3\\ 
\end{tabular}
\end{ruledtabular}
\end{table*}

\label{sectionI}

Mesas for Hall bars are defined using a solution of dilute phosphoric acid and citric acid. Following mesa definition, an Al etchant (Transene Type-D) is used to selectively etch the Al layer on the mesa. Following the wet Al-etch, a dielectric layer (hafnium oxide or aluminum oxide) is grown globally on top of the chip via atomic layer deposition (ALD), as illustrated in Fig.~\ref{fig_1}(a). Finally, the gate electrodes are patterned using Ti/Au deposition. The outer un-etched Al sections serve as ohmic contacts to the 2DEG.

Transport measurements are performed in dry dilution refrigerators with base mixing chamber temperatures of T$\sim$10~mK and up to a 6~T perpendicular magnetic field. We used standard AC lock-in techniques, applying a 10~nA AC current while simultaneously measuring the longitudinal voltage, $V_{xx}$, and the transverse voltage, $V_{xy}$, as a function of perpendicular magnetic field as shown in Fig.~\ref{fig_1}(b). 2DEG density is tuned by adjusting the DC voltage bias, $V_g$, on the top gate which is separated from the semiconductor by a dielectric layer.

\section{Results}
\subsection{Mobility versus 2DEG density}
\label{sectionA}

\begin{figure*}
\includegraphics[width=1.0\textwidth]{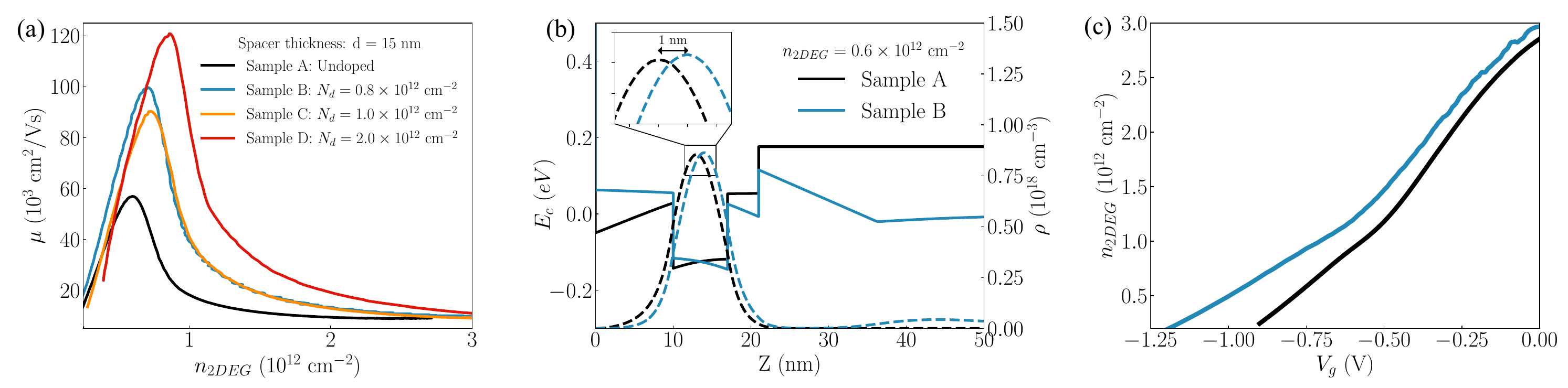}
\caption{\label{fig_2} (a) Mobility as a function of 2DEG density for Sample A to D with $N_d$=0, 0.8, 1.0, 2.0$\times10^{12}$~cm$^{-2}$ with spacer thickness $d$=15~nm. (b) Band diagram and charge distribution for Sample A and Sample~B calculated self-consistently at $n_{2DEG} = 0.6 \times 10^{12}$cm$^{-2}$. The solid line shows the conduction band edge relative to the Fermi energy ($E_F$ = 0~eV) as a function of the depth of the heterostructure, where $Z$=0~nm corresponds to the surface of the top barrier. Dashed lines display the spatial variation of the free charge distribution. The inset zooms to display the 1 nm shift of the center of the charge distribution due to the doping. (c) $n_{2DEG}$ vs. top gate voltage ($V_g$) for Sample A and B; note (c) and (a) share the same legend.
}
\end{figure*}
We began by investigating the impact of Si $\delta$-doping density on the relationship between mobility ($\mu$) and $n_{2DEG}$. $\mu$ vs. $n_{2DEG}$ for samples A to D with variable Si $\delta$-doping density but fixed 15~nm spacer thickness are shown in Fig.~\ref{fig_2}(a). $n_{2DEG}$ as function of $V_g$ is extracted from the Hall voltage at B$_{perp}$=0.5~Tesla. The black curve in Fig.~\ref{fig_2}(a) shows the mobility of sample A, the undoped structure similar in design to wafers used in previous experiments probing topological superconductivity~\cite{Dartiailh.2021an9, Banerjee.2022, Nichele.2017}. For Sample A, the peak mobility $\mu_{max}$ is equal to 57,000~cm$^2$/V~s at a 2DEG density of $n_{2DEG}=0.6\times10^{12}$~cm$^{-2}$. This peak mobility is typical for our heterostructure design without Si $\delta$-doping, and compares favorably to peak mobility reported previously in undoped structures ~\cite{Wickramasinghe.2018}. In different 2DEG density regimes distinct scattering mechanisms dominate, resulting in the observed non-monotonic dependence of $\mu$ on $n_{2DEG}$. The rapid increase in mobility with increased 2DEG density in the low density regime is characteristic of scattering from charged impurities in the vicinity of the 2DEG - presumably concentrated at the semiconductor-dielectric interface \cite{DasSarma.2015}. As the 2DEG density is increased, the mobility reaches a peak and then decreases. This sharp decrease begins at $n_{2DEG} \geq 0.6\times10^{12}$~cm$^{-2}$, corresponding to the density at which the chemical potential approaches the second subband; inter-subband scattering, alloy scattering, and interface roughness scattering decrease mobility at large 2DEG density ~\cite{DasSarma.2015, Davies.1997, Thomas.2018}.

Peak mobility in the doped samples increases significantly when compared to the undoped Sample A. The addition of a Si $\delta$-doping layer beneath the quantum well drastically alters the conduction band edge profile of the heterostructures. The simulated charge density and conduction band edge profile for the undoped Sample A and Sample B at $n_{2DEG}=0.6\times10^{12}$~cm$^{-2}$ are calculated using the NextNano$^3$ self-consistent Schrodinger-Poisson solver~\cite{Birner.2007}. Results are shown in Fig.~\ref{fig_2}(b). As indicated by the solid cyan line in Fig.~\ref{fig_2}(b), the ionized donors create an electric field in the quantum well pointing to the surface. This altered electric field profile compared to Sample A shifts the center of the 2DEG distribution 1nm away from the surface in Sample B as shown in the dashed lines in Fig.~\ref{fig_2}(b) and its inset. This 1nm spatial shift has substantial impact for shallow InAs 2DEG systems when Coulomb scattering from defects at or near the dielectric/semiconductor interface dominate ~\cite{Pauka.2020, Wickramasinghe.2018}. As discussed in Refs.~\cite{DasSarma.2015}, $\mu\propto d_{imp}^3$, where $\mu$ is the mobility of the 2DEG, and $d_{imp}$ is the distance between the remote 2D ionized impurities and the 2DEG. At $n_{2DEG}=0.6\times10^{12}$~cm$^{-2}$, the center of 2DEG distribution shifts from Z = 13 nm in Sample A to Z = 14 nm in Sample B; $\mu\propto d_{imp}^3$ implies a 24$\%$ increase of mobility due to the extra 1~nm separation in Sample B. We note at $n_{2DEG}=0.6\times10^{12}$~cm$^{-2}$ the mobility of Sample B increases approximately 80$\%$ compared to Sample A. In the simulation of Sample~B, a small population of electrons resides at the doped layer, as illustrated in Fig.~\ref{fig_2}(b). This population of electrons in the doping layer should enhance screening, resulting in additional increase in mobility. Screening by residual carriers in doping layers is known to improve mobility in the AlGaAs/GaAs 2DEG system \cite{manfra_2014, GARDNER201671}. Enhanced screening becomes more pronounced at higher doping density, as can be seen in Fig.~\ref{fig_2}(a). The slightly lower peak mobility of Sample C compared to Sample B may be attributed to two factors: 1) the relatively modest increase in Si doping density above Sample B; 2) slight fabrication run-to-run variations that impact the density of scattering centers at the semiconductor/dielectric interface. We also note that at high Si $\delta$-doping density $\geq$ 10$^{12}$cm$^{12}$ evidence of parallel conduction is seen in magnetotransport. This aspect will be covered more thoroughly in Section III B. 2DEG density vs gate voltage for Sample A and Sample B are shown in Fig.~\ref{fig_2}(c); a small increase in $V_g$=0~V density in evident in Sample B. In both devices a change in capacitance is visible below $V_g$$\approx$-0.5~V. This change in slope is attributed to the depopulation of the 2$^{nd}$ electric subband.

\begin{figure}
\includegraphics[width=0.48\textwidth]{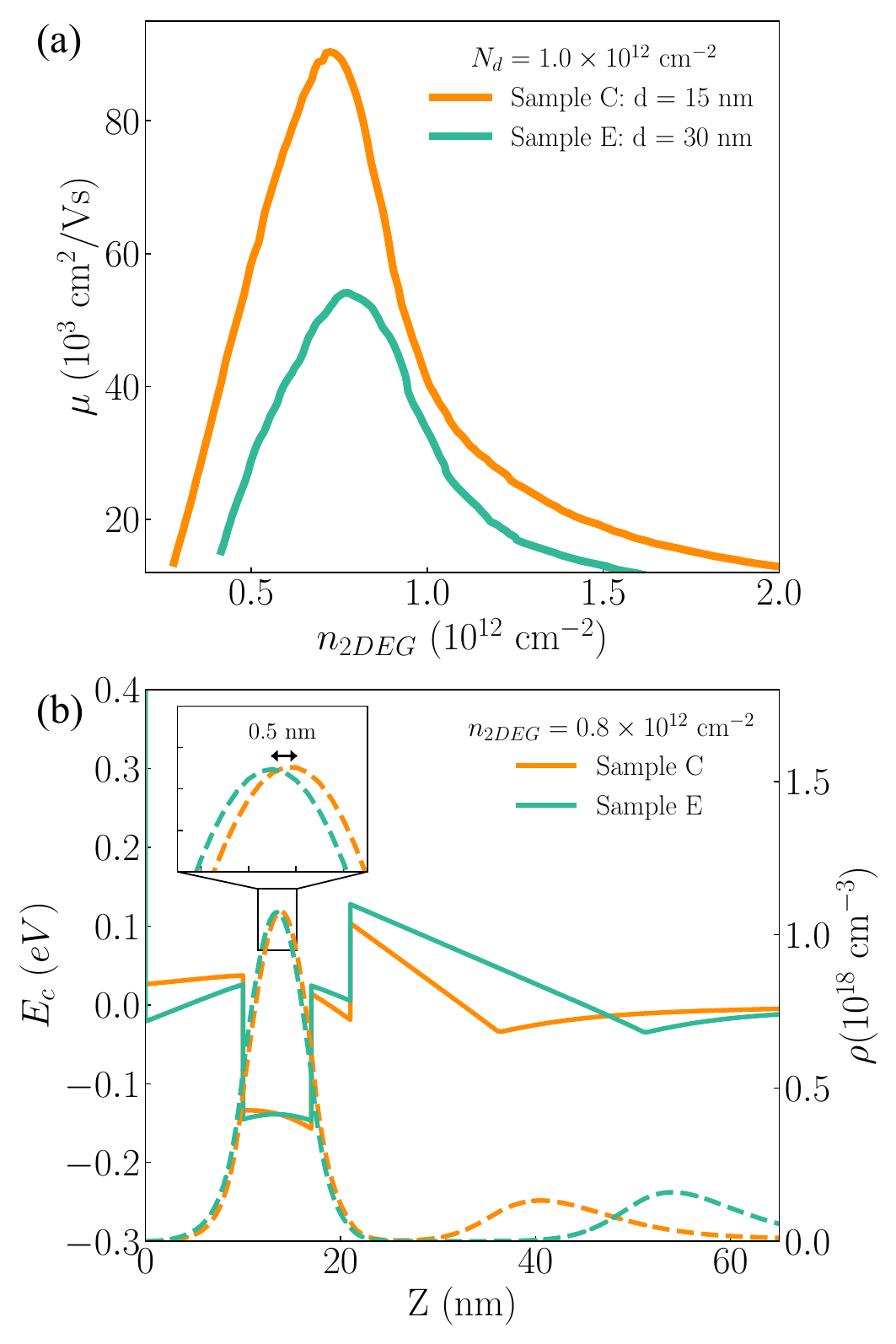}
\caption{\label{fig_3} (a) Mobility as a function of 2DEG density for Sample C ($d=15$nm) and sample E ($d=30$nm) with $N_d$ = 1.0$\times10^{12}$cm$^{-2}$. (b) Simulated conduction band edge diagram (solid lines) and charge distribution (dashed lines) for Sample C and Sample E at $n_{2DEG} = 0.8\times10^{12}$cm$^{-2}$. The inset zooms to the peak in charge density; a 0.5nm shift of the center of the charge distribution due to different spacer thicknesses is visible.
}
\end{figure}

The impact of spacer thickness is also studied. $\mu$ vs. $n_{2DEG}$ for Sample C ($d=15$~nm) and Sample E ($d=30$~nm) are shown in Fig.~\ref{fig_3}(a). Both samples have $N_d$=$1\times10^{12}$cm$^{-2}$. For all 2DEG densities studied here the mobility of Sample C with $d$=15nm is significantly higher than Sample E with $d
$=30nm. The peak mobility of Sample C is nearly twice that of Sample E at comparable densities. As shown in Table.~\ref{Table1}, such behavior is also observed when comparing Sample D and Sample F. This observation suggests that scattering from the ionized impurities in the doping layer is not a dominant scattering mechanism for spacer thickness d$\geq$15~nm. Moreover, the reduction of mobility with increased spacer thickness strongly suggests that the shift of charge distribution in the quantum well away from the dielectric/semiconductor interface is a dominant effect. The larger setback (30nm) results in a smaller shift relative to the 15nm spacer thickness. This is illustrated in Fig.~\ref{fig_3}(b) where the simulated spatial charge distributions of Sample C and Sample E are plotted at fixed 2DEG density. As shown in the insert in Fig.~\ref{fig_3}(b), the center of the charge distribution of Sample C is 0.5 nm further away from the surface compared to the situation of Sample E, resulting in a decrease in Coulombic scattering and an increase in the mobility of Sample C. 

\begin{figure}
\includegraphics[width=0.5\textwidth]{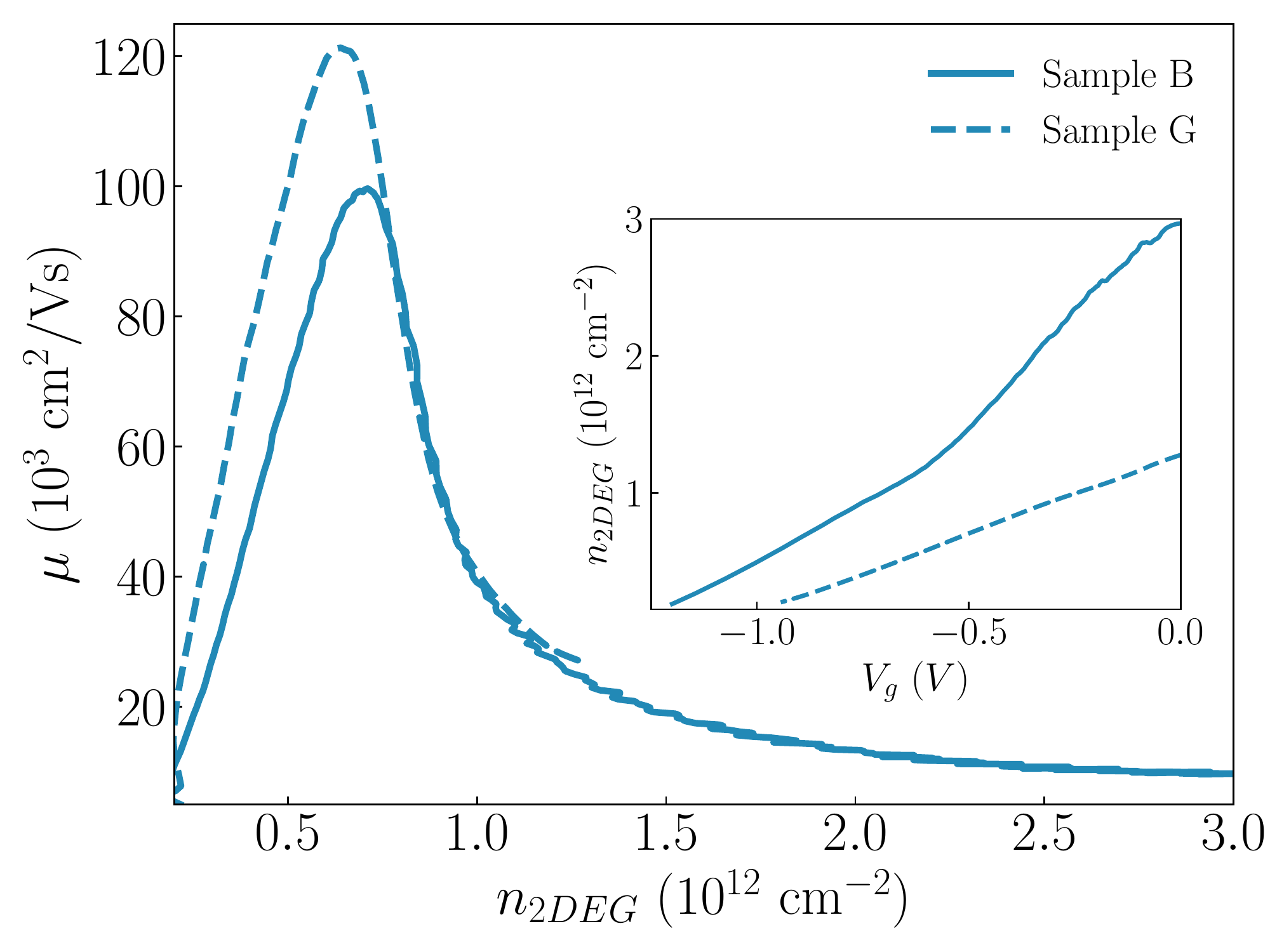}
\caption{\label{fig_4} Mobility as a function of 2DEG density for Sample B with 18nm HfO$_2$ dielectric layer (solid line) and Sample G with 19nm Al$_2$O$_3$ dielectric layer (dashed line). The inset shows that 2DEG density as a function of top gate voltage for Sample B (solid line) and sample G (dashed line).}

\end{figure}
As it appears that disorder at the dielectric-semiconductor interface is a primary limit to peak mobility in near-surface 2DEGs, we also test the role of the dielectric environment. We fabricated two chips using the semiconductor wafer with $N_d=0.8\times10^{12}$cm$^{-2}$ and $d$=15nm. This is the wafer used for Sample B. Sample B used 18nm HfO$_2$ as the gate dielectric while Sample G has 19nm of Al$_2$O$_3$. This change of dielectric is the only fabrication variation between the two samples. A comparison of mobility vs. 2DEG density for Sample B and Sample G is shown in Fig.~\ref{fig_4}; $n_{2DEG}$ vs. $V_g$ is shown in the inset. At zero gate voltage the 2DEG density for Sample G with 19nm Al$_2$O$_3$ is much lower than the density for Sample B with 18nm HfO$_2$, reflecting the difference in fixed charge density at the dielectric-semiconductor interface and difference in dielectric constant. A similar variation in 2DEG density in shallow InAs quantum wells depending on the details of surface preparation and choice of dielectric has been reported in Refs.~\cite{shabani.2021, Pauka.2020}. The peak mobility of Sample G with Al$_2$O$_3$ is $120,000$~cm$^2/$Vs; 20$\%$ higher than the peak mobility of Sample B with HfO$_2$ at similar 2DEG density. This increase in mobility is observed despite the reduction of dielectric constant from HfO$_2$~$\simeq19$ to Al$_2$O$_3$~$\simeq8$. The lower 2DEG density at $V_g$=0V combined with the higher peak mobility for Sample G suggest that the Al$_2$O$_3$/InGaAs interface has a lower interface state density than the HfO$_2$/InGaAs interface. Less free charge is transferred from the surface to the quantum well and consequently the fixed charged density remaining at the dielectric-semiconductor interface is reduced, leading to the improvement in peak mobility.
\begin{figure}
\includegraphics[width=0.5\textwidth]{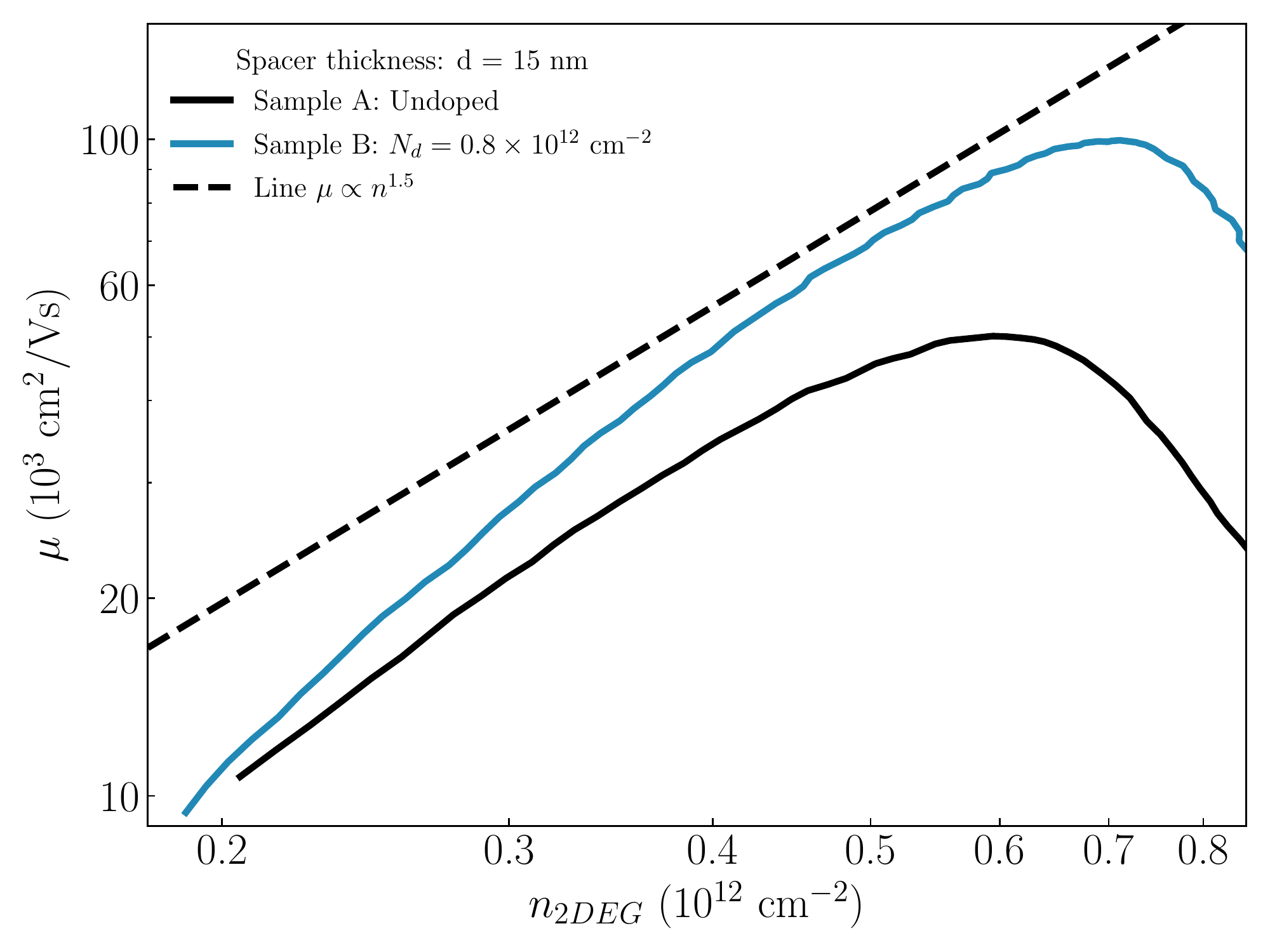}
\caption{\label{fig_5} $\mu$ vs. $n_{2DEG}$ in a log-log plot at low carrier density for Samples A and B. The black dashed line represents $\mu \propto n^{1.5}$. The dashed line is \textit{not} a fit to the data.}
\end{figure}

For low 2DEG density in the single subband limit where screened long-ranged Coulomb scattering dominates, mobility will have a power law dependence on $n_{2DEG}$, $\mu \propto n_{2DEG}^{\alpha}$ ~\cite{Stern.1967, Sarma.2013, DasSarma.2015}. The exponent $\alpha$ depends on the proximity of the charged Coulomb scattering centers to the quantum well and the strength of screening. Screening in our system may be parameterized by the Thomas-Fermi wavevector $q_{TF}=2m^*e^2/{\kappa}{\hbar}^2$. We estimated the Thomas-Fermi wavevector to be $q_{TF}$=0.065~nm$^{-1}$, where the effective electron mass (m$^*$=0.026m$_e$) and dielectric constant ($\kappa$=15) are approximated using bulk InAs values. Another important length scale is $1/k_F$ where $k_F$ is the Fermi wavevector of the 2DEG. At $n_{2DEG}\simeq$$7\times10^{11}$cm$^{-2}$, $1/k_F$=4.6~nm. The ratio $q_s=q_{TF}/2k_{F}$ sets an important scale; if $q_s\leq1$, as it is in the present case, the sample is in the weak screening limit \cite{Sarma.2013}. The product $k_{F}d$, where $d$ is the distance from the 2DEG to the ionized impurities, determines whether charged impurities are considered near or far. In the weak screening limit, the exponent $\alpha$ is predicted to asymptotically approach 1 for nearby 2D ionized impurities ($d_{imp}\leq2.3$~nm), while it is predicted to approach 1.5 for remote 2D ionized impurities ($d_{imp}>2.3$~nm) ~\cite{Sarma.2013}. Additionally, a 3D distribution of background charged impurities results in $\alpha \sim 1.5$~\cite{Sarma.2013}. Note that only in the strong screening limit ($q_s\gg 1$) with 3D impurities is the exponent expected to be less than 1. At very low 2DEG density, the electron system becomes localized and is not expected to follow power law behavior while at high 2DEG density near the transition to occupation of the 2$^{nd}$ electric subband other scattering mechanisms (e.g. intersubband scattering, alloy and interface roughness scattering) make significant contributions to scattering. $\mu$ vs. $n_{2DEG}$ is plotted on a log-log scale for Samples A and B in Fig.~\ref{fig_5}. The black dashed line is $\mu \propto n^{1.5}$; it is not a fit to the data but provides guide to the eye. The narrow density range with linear behavior precludes an accurate determination of a scaling exponent, but it is evident that the scaling is significantly greater than unity for both samples and approaches $\alpha\sim1.5$. This suggests that scattering is dominated by impurities at distances $d\geq 1/2k_F \geq 2.3$~nm, including the charge disorder at the dielectric/semiconductor interface ($d=10$nm) (samples A and B) and the intentional $\delta$-doping at $d=$15nm for sample B. It is interesting to note that the intentional introduction of impurities below the quantum well at $d=15$nm increases peak mobility rather than diminishes peak mobility. As discussed previously, the net effect of additional impurities is to reduce the potential fluctuations experienced by the 2DEG.

\subsection{Analysis of parallel conduction at high doping}
\label{sectionB}

\begin{figure}
\includegraphics[width=0.48\textwidth]{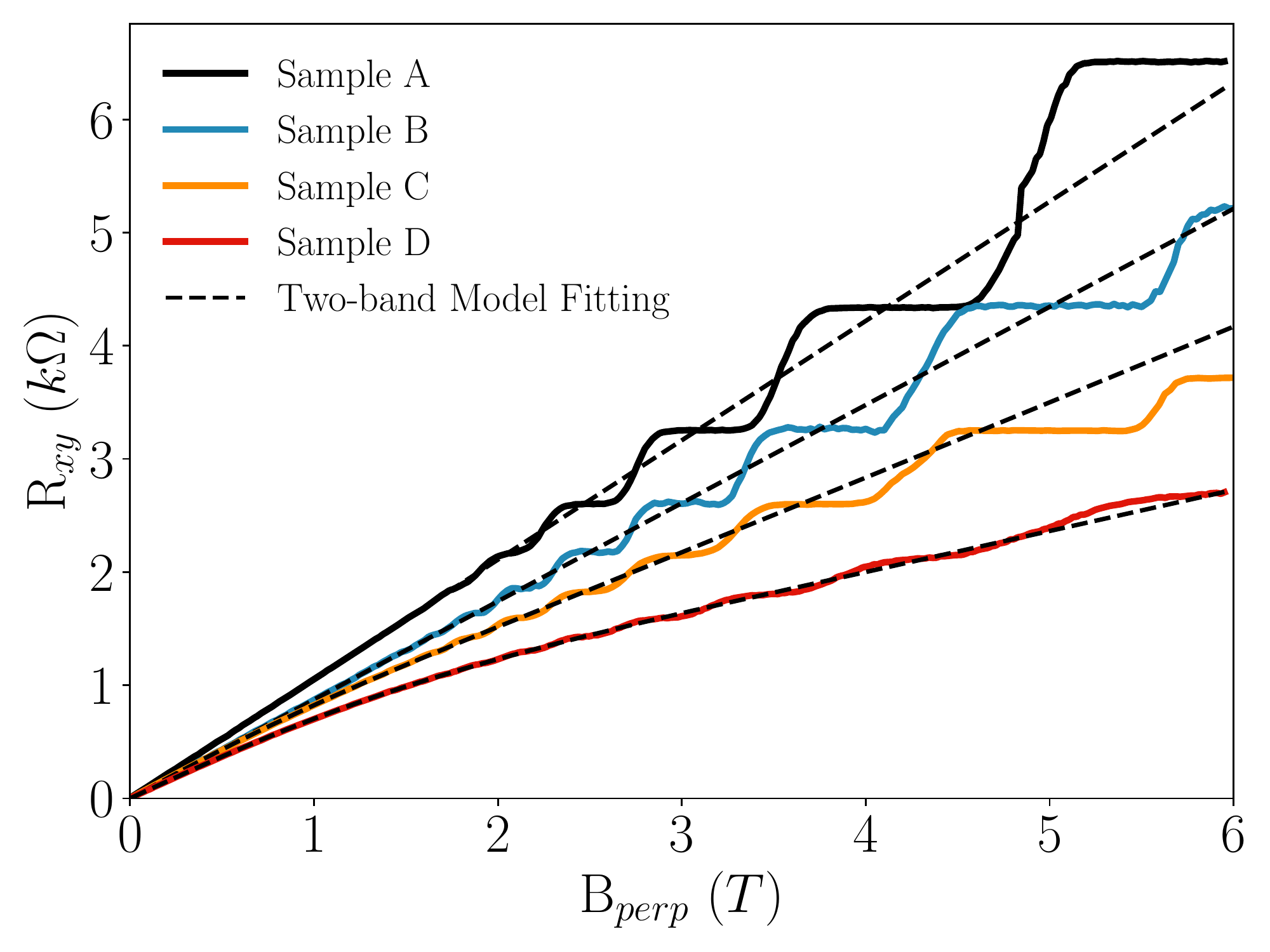}
\caption{\label{fig_6} Hall resistance, $R_{xy}$, as a function of the perpendicular magnetic field for Sample~A, B, C, and D at the 2DEG density that yields peak mobility for each sample. The fit using the two-band Drude model for each sample is shown by black dashed lines.}
\end{figure}

\begin{table}
\centering
\caption{\label{Table2}%
Carrier density ($n_{\parallel}$), and mobility ($\mu_{\parallel}$) in the parallel conduction channel extracted from the fit of the two-band Drude model.}
\begin{ruledtabular}
\begin{tabular}{>
    {\centering\arraybackslash}m{2cm}|>
    {\centering\arraybackslash}m{3cm}|>
    {\centering\arraybackslash}m{3cm}}
 Sample & $n_{\parallel}$~(10$^{12}$~cm$^{-2}$) & $\mu_{\parallel}$ ($10^3$~cm$^2$/Vs)\\
 \hline
 A  & 0    & 0 \\ 
 B  & 0    & 0 \\ 
 C  & 0.23 & 4.8  \\ 
 D  & 0.75 & 2.1\\ 
 E  & 0.55 & 1.2\\ 
 F  & 1.11 & 1.4 \\ 
\end{tabular}
\end{ruledtabular}
\end{table}

As we increase the Si $\delta$-doping density, the minimum in the conduction band edge will eventually dip below the Fermi level at the doped layer, forming an unintentional parallel conduction channel. This situation is undesirable and it is important to understand the onset of parallel conduction and its impact on transport measurements. The carrier density in any unintentional parallel channel can be estimated using the two-band Drude model to fit the Hall resistance~\cite{Chambers.1952, Peters.2017}. 
\begin{equation}
     \begin{split}
         R_{xy} =&  B\gamma \gamma_{||}(e n_{2DEG}\mu^2 \gamma_{||}+e n_{||}\mu_{||}^2 \gamma)\\
     &\big([e n_{2DEG}\mu \gamma_{||}+e n_{||}\mu_{||} \gamma]^2 +\\ 
     &[e n_{2DEG}\mu^2 B \gamma_{||}+ e n_{||}\mu_{||}^2 B \gamma]^2\big)^{-1},
     \end{split}
\end{equation}
with $\gamma = 1 + \mu^2 B^2$ and $\gamma_{||} = 1+\mu_{||}^2 B^2$. 
There are four parameters in the two-band Drude model: the 2DEG density ($n_{2DEG}$), the mobility ($\mu$) of the 2DEG, the carrier density in the parallel channel ($n_{||}$), and the mobility of carriers in the parallel channel ($\mu_{\parallel}$). 

In Fig.~\ref{fig_6}(a), the Hall resistance of Samples A, B, C, and D is presented. In order to reduce the number of free fitting parameters, we estimate the mobility of the 2DEG from the measured value of the $B=0$ Tesla resistivity and the value of $R_{xy}$ at $B=0.5$ Tesla. As we expect the 2DEG to dominate the conductivity, this is a reasonable approximation. We then fit $R_{xy}$ to the two-band model over the full field range $B \leq 6$ Tesla. The extracted parameters for the parallel channel in Samples A through F are summarized in Table.~\ref{Table2}. As expected, the parallel channel density increases with increasing doping density and spacer thickness. Since the doping layer is highly disordered, the mobility of carriers in the parallel channel is very low. The low mobility of carriers in the parallel channel, compared to that the principal 2DEG, indicates transport is dominated by the 2DEG. The heterostructure design with $N_d=0.8\times10^{12}$cm$^{-2}$ at $d=15$~nm setback appears to be nearly optimal, as it shows a peak mobility exceeding 100,000~cm$^2$/Vs without unintentional parallel conduction.

\subsection{High magnetic field measurements}
\label{sectionC}

\begin{figure*}
\includegraphics[width=1.0\textwidth]{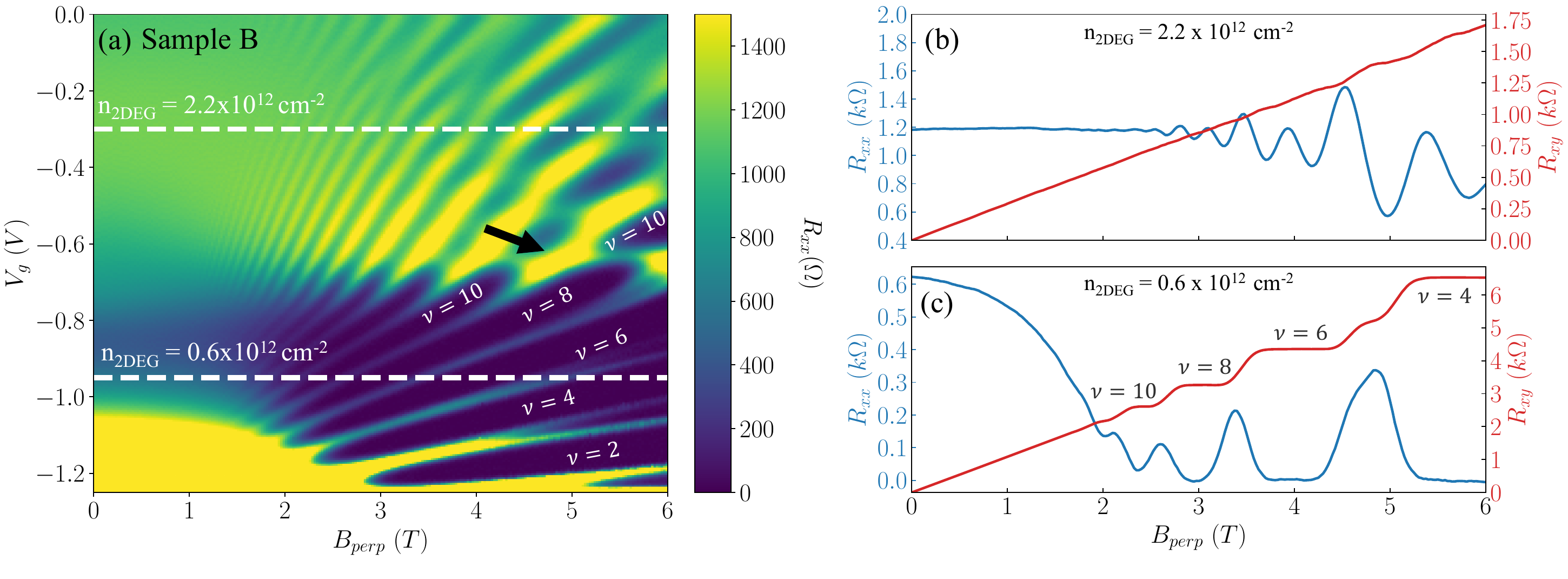}
\caption{\label{fig_7}(a) $R_{xx}$ as a function of the top gate voltage $V_g$ and the perpendicular magnetic field $B_{perp}$. The black arrow indicates the location of a Landau level crossing. Two white dashed lines indicate the position of line cuts shown in (b) and (c). Filling factors of the integer quantum Hall states are indicated.} 
\end{figure*}

\begin{figure}
\includegraphics[width=0.48\textwidth]{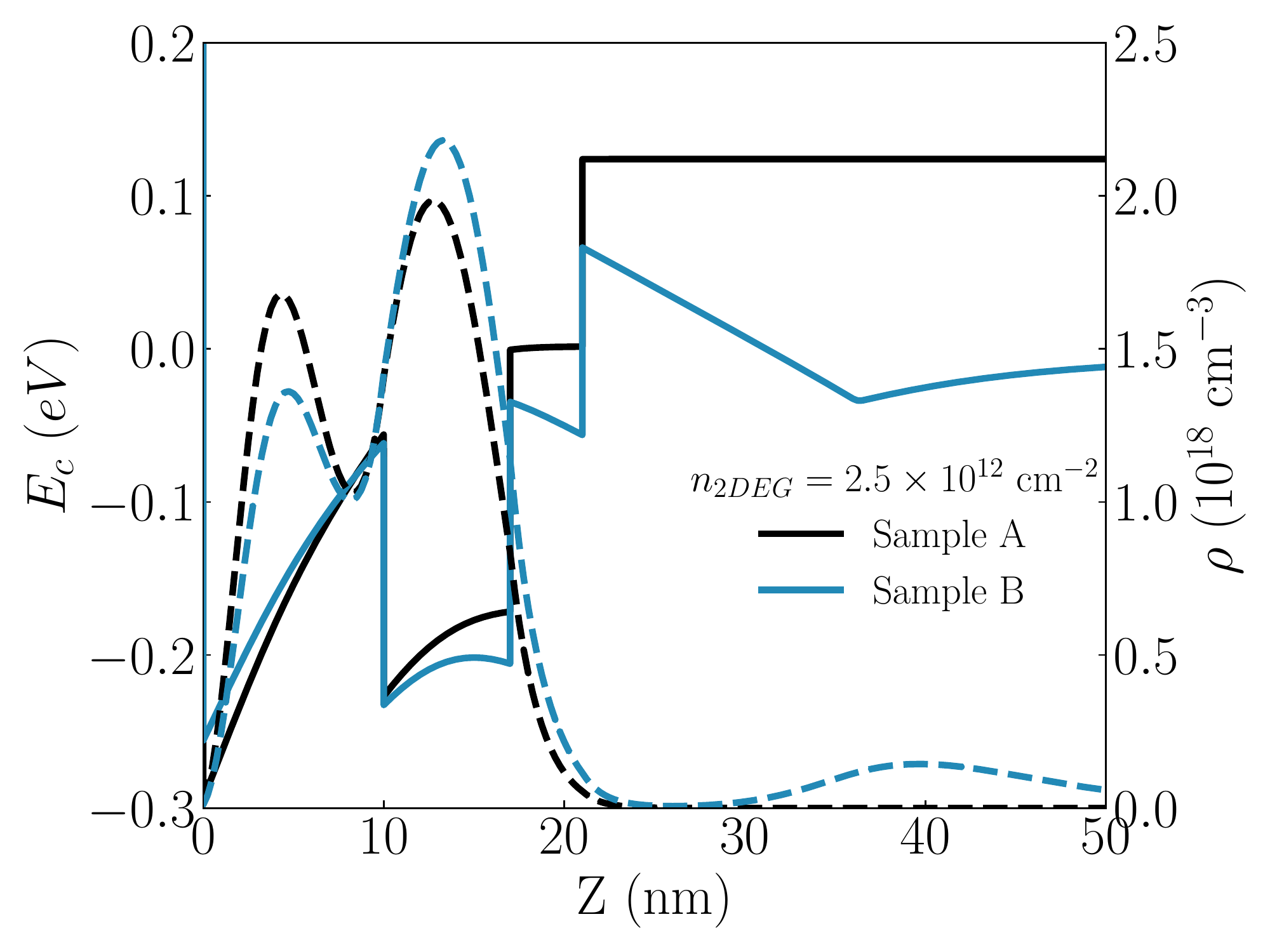}
\caption{\label{fig_8} Band structure and charge distribution for Sample~A and Sample~B from self-consistent Schrodinger-Poisson calculations at $n_{2DEG}=2.5\times10^{12}$~cm$^2$ 
}
\end{figure}

Fig.~\ref{fig_7}(a) illustrates the longitudinal resistance of Sample~B as a function of the top gate voltage and the perpendicular magnetic field. The filling factor $\nu$ is defined by $\nu=n_{2DEG}\phi_{0}/B_{perp}$ where $\phi_{0}=h/e$ is the magnetic flux quantum. The black arrow in Fig.~\ref{fig_7}(a) identifies the location of a Landau level crossing in the vicinity of $\nu = 10$. The Landau level crossing suggests that the second subband is occupied at $n_{2DEG}=1.1\times10^{12}$~cm$^{-2}$ ($V_g=-0.65$~V)~\cite{Ellenberger.2006, Zhang.2005, Yuan.2020}. $R_{xx}$ and the transverse resistance $R_{xy}$ for $n_{2DEG}=2.2\times10^{12}$~cm$^{-2}$ are shown in Fig.~\ref{fig_7}(b). Two distinct sets of Shubnikov de Haas (SdH) oscillations are visible in $R_{xx}$ in Fig.~\ref{fig_7}(b), confirming second subband occupation. In Fig.~\ref{fig_7}(c) the system is in the single subband regime, and when the perpendicular magnetic field is greater than 3~T, quantized Hall states are visible in $R_{xx}$ and $R_{xy}$. The integer quantum Hall states seen in Sample~B further confirm the absence of parallel channel in the structure and the material's high quality. As demonstrated by a self-consistent simulation in Fig.~\ref{fig_8}, the second subband has significant weight in the top InGaAs barrier, which is consistent with our experimental results: mobility is decreased significantly when the $2^{nd}$ subband is occupied and the gate first depletes the second subband, as seen in the Fig.~\ref{fig_2} (c) as a change in slope (equivalently capacitance) for $n_{2DEG}$ vs. $V_g$.

\subsection{Rashba spin-orbit coupling}
\label{sectionD}

\begin{figure}
\includegraphics[width=0.48\textwidth]{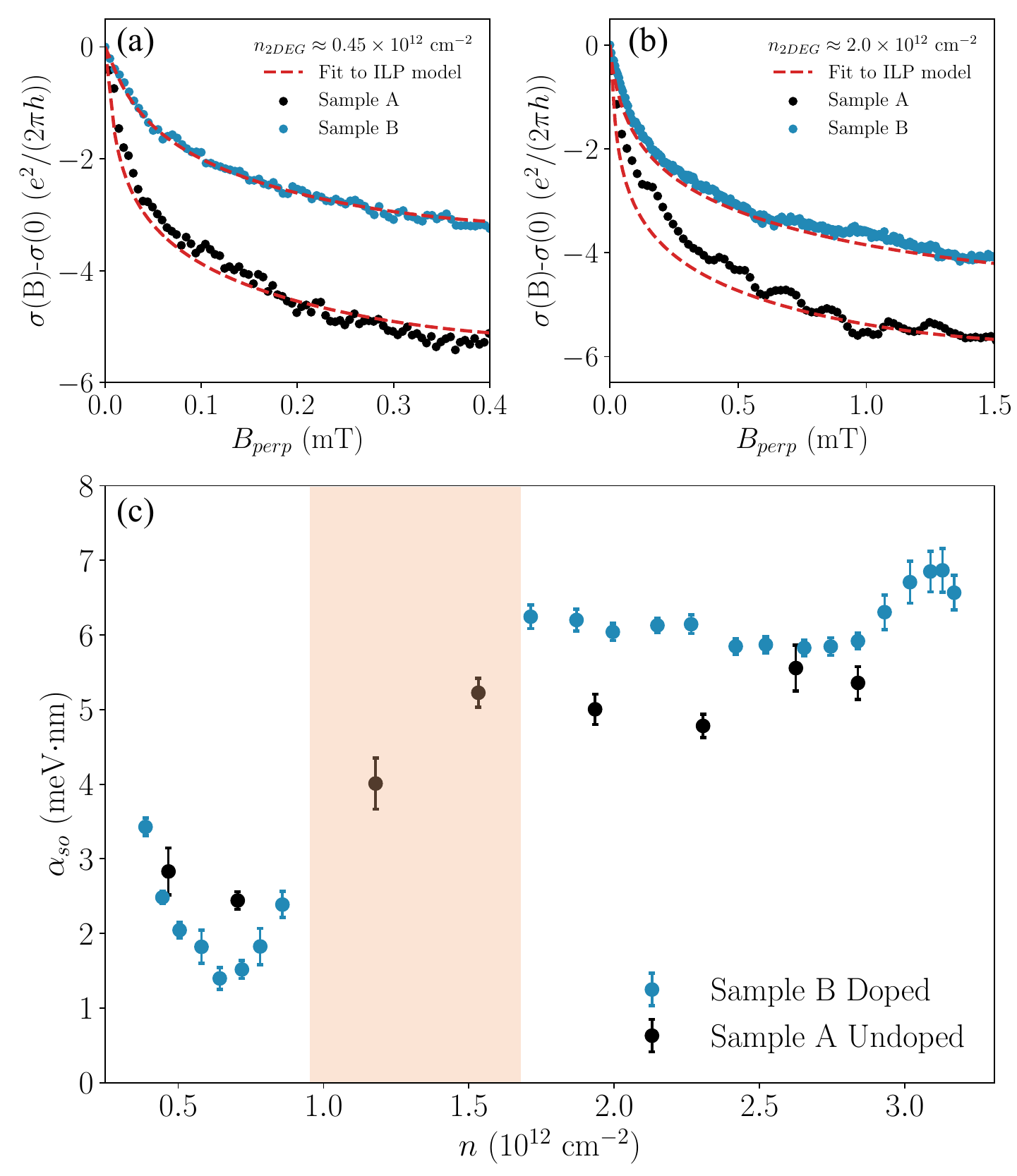}
\caption{\label{fig_9} Low field magnetoconductivity of Sample~A and Sample~B for densities $n_{2DEG}=0.4\times10^{12}$~cm$^{-2}$ in (a) and $n_{2DEG} =2\times10^{12}$~cm$^{-2}$ in (b). Fits to the ILP model are indicated by the red dashed lines. (c) The Rashba spin-orbit coupling strength, $\alpha_{so}$~(meV$\cdot$nm), for Sample~A and Sample~B, as a function of the 2DEG density. Values for $\alpha_{so}$ were extracted from the fits to the ILP model. The shaded region for Sample~B designates $l_e\geq l_{R-so}$, where model cannot be applied reliably.}
\end{figure}

We explored the impact of heterostructure design on spin-orbit coupling in our devices. Rashba spin-orbit coupling is assessed by analysis of weak antilocatization 
(WAL) behavior in low-field magnetoconductivity measurements.  Large Rashba SOC promotes topological superconductivity; we need to quantify if the enhanced mobility in our $\delta$-doped samples comes at the expense of diminished Rashba coupling associated with modifications of the electric field profile in the growth direction. The Rashba coupling strength is quantified by Rashba paramter $\alpha_{so} = e\alpha_0\langle E \rangle$, where $e$ is the electronic charge, $\alpha_0$ is the intrinsic Rashba parameter related to band properties of the host materials, and $\langle E \rangle$ is the average electric field in the region where 2DEG resides. The measurement is carried out in a small perpendicular magnetic field range around $B_{perp}=0$ Tesla. Fig.~\ref{fig_9}(a) and (b) illustrate the symmetrized conductivity as a function of the perpendicular magnetic field for both Sample~A and Sample~B, with $n_{2DEG}=0.45\times10^{12}$~cm$^{-2}$ in the single subband region (a) and $n_{2DEG}=2.0\times10^{12}$~cm$^{-2}$ in the two-subband region (b). The red dashed lines in Fig.~\ref{fig_9}(a) and (b) are fits to the model of Iordanski, Lyanda-Geller, and Pikus (ILP)~\cite{Knap.1996}. 

\begin{equation}
    \begin{split}
        \Delta \sigma(B) =& \frac{-e^2}{4\pi^2\hbar}\bigg(\frac{1}{a_0} + \frac{2a_0+1+\frac{B_{so}}{B}}{a_1(a_0+\frac{B_{so}}{B})-\frac{2B^{'}_{so}}{B}}-\\
         & \sum_{n=0}^{\infty}[\frac{3}{n}-\\
         &\frac{3a^2_n+2a_n\frac{B_{so}}{B}-1-2(2n+1)\frac{B^{'}_{so}}{B}}{(a_n+\frac{B_{so}}{B})a_{n-1}a_{n+1}-2\frac{B^{'}_{so}}{B}[(2n+1)a_n-1]}]\\
         &+2\ln{\frac{B_{tr}}{B}}+\Psi(\frac{1}{2}+\frac{B_{\phi}}{B})+3C\bigg),
    \end{split}
\end{equation}
with $\Delta \sigma(B) = \sigma(B)-\sigma(0)$, $B_{so} = \frac{\hbar}{4el^2_{R-so}}+\frac{\hbar}{4el^2_{D-so}}$, $B^{'}_{so} = \frac{\hbar}{4el^2_{R-so}}$, $B_{\phi} = \frac{\hbar}{4el^2_{\phi}}$, $B_{tr} = \frac{\hbar}{2el^2_{e}}$, $a_n = n+\frac{1}{2} + \frac{B_{\phi}}{B} + \frac{B_{so}}{B}$, $C$ is Euler's constant and $\Psi$ the Digamma function. This expression for the magnetoconductivity depends on four parameters: the spin relaxation length due to Rashba SOC ($l_{R-so}$), the spin relaxiation length due to Dresselhaus SOC ($l_{D-so}$), the quantum phase coherence length ($l_{\phi}$), and the mean free path ($l_{e}$). The Rashba parameter is given by $\alpha_{so} = \frac{\hbar^2}{2l_{R-so}m^*}$.

To apply ILP model reliably, length scales need to follow a hierarchy: ($l_{\phi}, l_{R-so}$, and $l_{D-so})\geq l_{e}$. The mean free path $l_{e}$ is extracted from the mobility vs. density data examined in Section~\ref{sectionA}. In the two subband regime, we fit the data in a range of perpendicular magnetic field $|B_{perp}|<0.5~B_{tr}$. In the single subband regime, we used the range $|B_{perp}|<0.7~B_{tr}$ in order to improve the reliability of the fit. The Rashba parameter is estimated through $\alpha_{so} = \frac{\hbar^2}{2l_{R-so}m^*}$, in which the effective mass $m^*$ is estimated with the bulk InAs electron mass $0.023~m_{e}$, where $m_{e}$ is the electron mass in vacuum. The Rashba parameter extracted from the ILP model is plotted as a function of the 2DEG density in Fig.~\ref{fig_9}(c) for both Sample~A and Sample~B.

In Fig.~\ref{fig_9}(c), $\alpha_{so}$ exhibits a non-monotonic dependence on 2DEG density for Sample~B in the single subband regime ($n_{2DEG}<1.1\times 10^{12}$~cm$^{-2}$). The non-monotonic behavior can be understood from the fact that both the surface charge and the Si $\delta$-doping layer determine the electrical field profile across the the quantum well. As the gate voltage is tuned to change the 2DEG density, the electric field at the location of the 2DEG goes through a local minimum. Either increasing or decreasing the gate voltage from this point will increase the asymmetry of the potential in the quantum well, resulting in an increase in the electric field. When $n_{2DEG}\approx0.6\times10^{12}$~cm$^{-2}$, as seen in Fig. \ref{fig_9}(c), there is a local minimum of the Rashba parameter which suggests that this is the point where the quantum well is most symmetric. As the 2DEG is increased from this value, $\alpha_{so}$ increases.

In the two-subband regime $n_{2DEG}>1.1\times10^{12}$~cm$^{-2}$, the Rashba parameter saturates at approximately $5$~meV$\cdot$nm~for Sample~A and at approximately $6$~meV$\cdot$nm~for Sample~B. At high 2DEG density, the $2^{nd}$ subband is increasingly populated. The $2^{nd}$ subband has significant weight in the InGaAs top barrier. Given that the intrinsic Rashba parameter for InGaAs ($36.9~\AA^2$) is substantially smaller than for InAs ($117~\AA^2$) in the K$\cdot$P model~\cite{Knap.1996}, the contribution from the second subband tends to decrease the effective Rashba parameter. This mechanism might explain the saturation of extracted Rashba parameter in the two-subband regime for both Sample~A and Sample~B. The slightly larger Rashba value in Sample~B at high 2DEG density may be due to the Si doping layer's effect on the charge distribution. Fig.~\ref{fig_8} illustrates the band structure and charge distribution of Samples A and B obtained using self-consistent Schrodinger-Poisson calculations in the two-subband regime with $n_{2DEG}=2.5\times 10^{12}$~cm$^{-2}$. The dashed lines in Fig.~\ref{fig_8} clearly show that Sample~B has more carriers in the InAs quantum well and fewer carriers in the top barrier than Sample~A at the same total carrier density, which could explain Sample~B's slightly larger Rashba parameter in the two-subband regime.

The impact of Si $\delta$-doping on $\alpha_{so}$ evidently depends on the density of the 2DEG. At low $n_{2DEG}$ it appears that $\alpha_{so}$ may reach a deeper minimum in sample~B around $n_{2DEG} \simeq 6 \times 10^{11}$cm$^{-2}$ as the quantum well potential profile is more symmetric than in sample A, while at higher densities Sample~B reaches slightly larger values of $\alpha_{so}$ than Sample A, presumably due to the weight of the $2^{nd}$ in the top barrier as discussed previously. Overall, the differences in $\alpha_{so}$ at any particular density are not very large when considering the enhanced mobility in Sample B. Rashba parameters, $\alpha_{so}$, reported in the literature for similar shallow InGaAs/InAs 2DEGs systems at approximately similar 2DEG densities in the single subband regime are around 1.5 to 2.5meV$\cdot$nm ~\cite{Wickramasinghe.2018, Farzaneh.2022}, albeit at lower 2DEG mobility. Witt et. al. \cite{Witt.2021} reported around 1.5~meV$\cdot$nm on shallow 2DEGs with InAlAs top barrier. The values extracted in this study are comparable to those mentioned above. Our study illustrates the interplay of doping, 2DEG density, and electric field profile in determining the magnitude of $\alpha_{so}$ for different regimes of operation. 

\subsection{Induced Superconductivity}
\label{sectionE}
\begin{figure}
\includegraphics[width=0.5\textwidth]{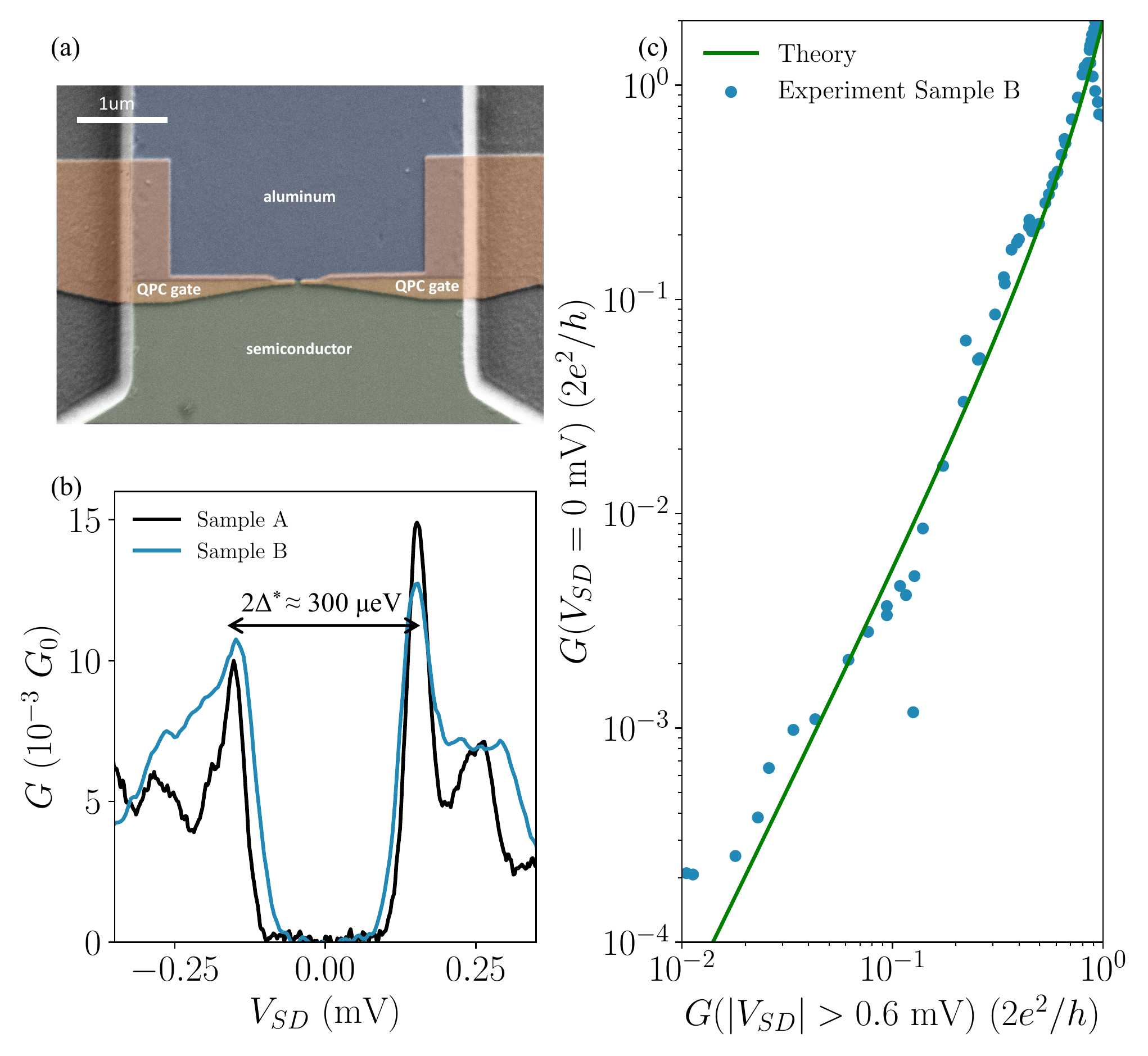}
\caption{\label{fig_10} (a) False-color scanning electron microscope (SEM) image of superconductor-QPC-semiconductor device. The Al layer is wet-etched in the green shaded region and is untouched in blue shaded region. An 18~nm Hafnium Oxide layer (not shown) separates the QPC gates and the heterostructure. (b) Conductance as a function of the DC bias voltage when QPC gates are at the tunneling region for Sample~A and Sample~B, both at zero magnetic field. (c) Differential conductance at zero source-drain bias as a function of the averaged differential conductance at finite source-drain bias for Sample~B. The solid green line is a theoretical prediction for the conductance of the perfectly transparent junction without free parameters.   
}
\end{figure}

In this section, we explore induced superconductivity in Sample~B and compare the results to those obtained in Sample~A. One of the advantages of shallow InAs quantum wells is the high transparency of the epitaxial superconductor-semiconductor interface, resulting in a hard induced superconducting gap~\cite{Kjaergaard.2016, Drachmann.2017, Chang.2015}. To investigate the induced superconducting gap, we fabricated superconductor-quantum point contact-semiconductor devices (SQPCN) for Samples~A and B, as shown in Fig.~\ref{fig_10}(a). This device is used to perform tunneling spectroscopy on the InAs 2DEG proximitized by aluminum. The tunnel barrier is regulated by the QPC gate voltage, $V_{QPC}$. The junction's differential conductance, $G$, is measured as a function of the source drain bias, $V _{SD}$.

The differential conductance $G$ as a function of the source drain bias $V_{SD}$ is shown in Fig.~\ref{fig_10}(b), where the QPC has been biased into the tunneling regime for both Samples~A and B. The differential conductance reflects the local density of states, and induced superconducting gap may be directly observed~\cite{Kjaergaard.2016, Beenakker.1992}. The extracted gaps based on the coherence peak-to-peak separations are approximately $150~$\textmu{}eV for both Samples~A and B, indicating that the Si doping layer has a minimal effect on the magnitude of the induced gap. In both samples, hard induced gaps can be seen in Fig.~\ref{fig_10}(b). To further characterize the induced gap, we performed differential conductance measurements in which the transmission of the QPC barrier is varied. The QPC transmission is parameterized by the value of above gap conductance. The sub-gap conductance ($G(V_{SD}=0$~V$)$) as a function of the above gap conductance ($G(|V_{SD}|>0.6$~mV$)$) is plotted on a log-log scale in Fig.~\ref{fig_10}(c). The data in Fig.~\ref{fig_10}(c) are compared to theoretical predictions for a perfect super-semiconductor interface~\cite{Beenakker.1992},
\begin{equation}
    G_{S} = 2G_{0}\frac{G_{N}^2}{(2G_{0}-G_{N})^2}
\end{equation}
with no fit parameters. Here, $G_S$ denotes the sub-gap conductance (measured at zero source drain bias), $G_N$ represents the above-gap conductance (measured at a high source drain bias, $|V_{SD}|>0.6$~mV in our experiment), and $G_0$ represents the conductance quantum. Good agreement between the experiment and theory over four orders of magnitude is found, indicating that the superconductor-semiconductor interface in Sample~B is transparent.

\section{Conclusion}
In summary, we have systematically studied the impact of modulation-doping in shallow 2DEGs in InGaAs/InAs heterostructures coupled to epitaxial aluminum. We observed peak mobility exceeding 100,000~cm$^2$/V~s in a shallow InAs quantum well when Si $\delta$-doping with density of $0.8\times10^{12}$~cm$^{-2}$ is placed 15~nm below the quantum well. Increasing doping density or changing the setback reduces mobility and/or induces parallel conduction. We compared the Rashba SOC parameter $\alpha_{so}$ as a function of 2DEG density in an updoped and optimally doped sample. Enhancement of mobility associated with $\delta$-doping can be realized without significant reduction in spin-orbit coupling, a necessary component for generation of topological phases. Our findings suggest additional strategies for optimization of complex superconductor-semiconductor heterostructures in which disorder must be reduced to promote strong signatures of topological properties.

\section{Acknowledgements}
We thank N. Hartman for preliminary low temperature measurements made at an early stage of this project. This work was supported by Microsoft Quantum.

\bibliography{citedpaper}

\end{document}